\documentclass[11pt]{article}
\usepackage[margin=1in]{geometry}
\usepackage{amsmath,amssymb}
\usepackage{booktabs}
\usepackage{graphicx}
\usepackage{hyperref}
\usepackage[numbers,sort&compress]{natbib}

\usepackage{multirow}

\title{Physical Law Ecology: mapping multi-mechanism ecologies as the zeroth step of data-driven scientific discovery}
\author{Xiongheng Bian$^{1,*}$,Xiangyu Cui$^{1}$,Ma Feng$^{2}$, Xiaoyan Shen$^{1,*}$\\[4pt]
\small $^1$School of Information Science and Technology, Nantong University, Nantong 226019, China\\
\small $^2$School of Artificial Intelligence and Computer Science, Nantong University, Nantong 226019, China\\
\small $^*$Corresponding authors: X. Bian (BXH@ntu.edu.cn); X. Shen}
\date{}

\begin{document}
\maketitle

\begin{abstract}

Every data-driven equation discovery method assumes (implicitly and without verification) that the target system obeys a single governing law ($K{=}1$). Here we show that this assumption is the primary bottleneck limiting scientific discovery in multi-mechanism systems, and introduce Physical Law Ecology, a framework that makes $K^*$ (the number of coexisting independent mechanisms) itself the first quantity to be determined from data. The framework automatically mines a pool of topologically distinct candidate equations, constructs a continuous dominance weight field across parameter space, and discovers analytic evolution laws governing mechanism succession---with optional monotonicity constraints encoding irreversible physics. Across four unrelated systems (elastomer mechanics, pool boiling, galactic dynamics, and droplet evaporation), BIC consistently identifies $K^*{=}3$ independent governing topologies. Applied to 163 SPARC galaxies (3,269 spatially resolved measurements), the framework autonomously recovers three gravitational laws whose coexistence provides evidence against the single-universal-acceleration hypothesis of MOND ($p<10^{-34}$). In engineering applications, multi-law weighted prediction reduces error by 67-72\% over single-equation baselines while retaining full interpretability. By establishing the determination of $K^*$ as the zeroth step of scientific discovery-prior to and independent of equation search---this work opens a direction orthogonal to existing symbolic regression: not finding better equations, but mapping the ecology of mechanisms that govern complex systems.
\end{abstract}

\section*{Introduction}

The development of modern physics has been guided by the pursuit of single-law paradigms. From Newton's law of gravitation and Maxwell's electromagnetic equations to classical thermodynamics, scientific inquiry has followed the core assumption that a single physical phenomenon corresponds to a unique universal governing equation, building an elegant and unified theoretical edifice. In recent years, symbolic regression (SR)\cite{schmidt2009distilling,cranmer2023pysr,lacava2021contemporary} has emerged as a powerful alternative to traditional mechanistic derivation\cite{brunton2016sindy,udrescu2020aifeynman,karniadakis2021physics}, offering interpretable, model-free, automatic extraction of explicit physical equations. However, existing SR methods inherit the single-law assumption wholesale: they fit one optimal explicit equation to a given dataset, presuming the system obeys a control law of fixed topological structure throughout. This assumption holds for simple homogeneous systems but completely fails for real complex physical systems. The vast majority of natural physical systems exhibit multi-mechanism coupling, temporally induced phase transitions, and qualitative regime switching---different operating conditions, spatial locations, and evolutionary stages correspond to fundamentally different governing mechanisms that cannot be captured by tuning parameters of a single equation.

A more fundamental blind spot exists: all current methods---whether classical mechanistic derivation or data-driven symbolic regression---default to $K=1$ (the system obeys a single governing equation) and proceed directly to searching for the optimal equation. Yet for unknown complex systems, the question that must be answered before searching for equations is: \textbf{How many topologically independent governing mechanisms exist in the system?} The absence of this ``zeroth step'' means that existing methods facing multi-mechanism systems can only find local optima and cannot map the complete mechanism ecology.

To address this gap, we propose Physical Law Ecology, a framework that treats $K^*$---the true number of mechanisms in a system---as the primary quantity to be discovered rather than a human-preset constant. The framework draws an analogy between complex multi-mechanism physical systems and natural ecosystems: governing equations of different topological structures act as ecological species occupying dedicated niches in parameter space, competing dynamically to achieve dominant control at different times and spatial locations, forming a dynamically balanced coexistence of laws. We present the complete algorithmic architecture, theoretical foundations, and cross-domain empirical validation.

Several existing lines of work address systems with multiple regimes but differ from our framework in fundamental ways. Switching system identification methods (e.g., hybrid SINDy variants\cite{brunton2016sindy}) segment data into discrete regimes and fit separate models per segment, but require the number of regimes and their boundaries as input, producing hard partitions rather than continuous weight fields. Bayesian model averaging\cite{mangan2017model} computes posterior probabilities over a pre-specified model set but does not discover new equation topologies from data. Multi-modal symbolic regression approaches\cite{cranmer2020discovering} learn latent representations that inform equation search but still output a single best-fit expression. Our framework is distinguished by three properties simultaneously: (i) \emph{automaticity}---$K^*$ is determined from data via BIC without human specification of mechanism counts or partition boundaries; (ii) \emph{continuity}---dominance weights vary smoothly across parameter space, capturing gradual mechanism transitions rather than hard switches; (iii) \emph{full interpretability}---every output is a closed-form analytic equation, not a black-box model.

\section*{Results}

\subsection*{1\quad Framework overview}

\begin{figure}[htbp]
  \centering
  \includegraphics[width=\textwidth]{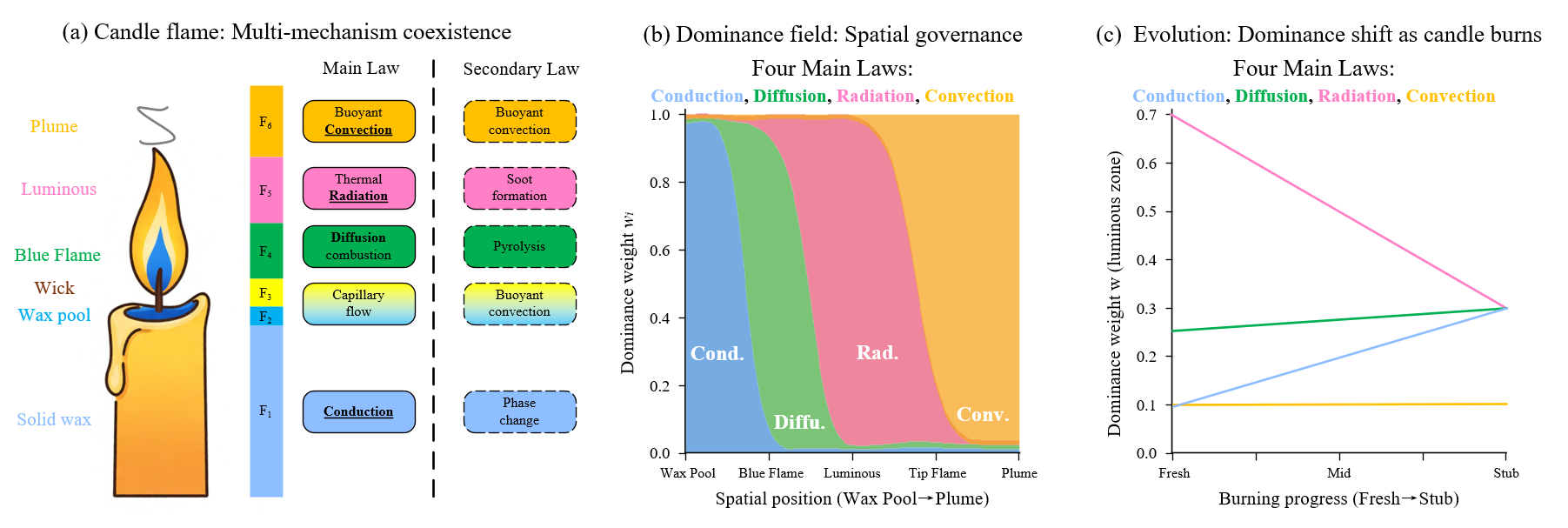}
  \caption{\textbf{Candle combustion: an intuitive example of Physical Law Ecology.} (a) Different spatial regions of a candle flame are governed by different physical mechanisms---from heat conduction in the wax pool to diffusion combustion in the blue zone, radiative heat transfer in the luminous zone, and buoyant convection at the tip. (b) The dominance field quantifies the governance weight of each law at every spatial location. (c) As the candle burns shorter, dominance weights at a given location evolve: radiation weakens while conduction strengthens. The Physical Law Ecology framework automatically discovers this spatial multi-law coexistence pattern and its conditional evolution from data.}
  \label{fig:candle}
\end{figure}

We first use candle combustion as an intuitive example to illustrate the core idea (Fig.~\ref{fig:candle}). A candle flame appears simple but is in fact governed by multiple physical mechanisms in spatially layered fashion: the wax pool region is governed by the heat conduction equation, the blue inner flame by diffusion combustion, the bright outer flame by radiative heat transfer, and the top plume by buoyant convection (Fig.~\ref{fig:candle}a). Applying conventional symbolic regression ($K=1$) to this system yields only a single ``average'' equation, discarding all spatial heterogeneity. Figure~\ref{fig:candle}b shows the dominance weight field output by our framework---the governance weights of the four equation classes vary continuously with spatial position, precisely quantifying ``who dominates where.'' More importantly, as the candle burns shorter, the weights at a given location undergo systematic evolution (Fig.~\ref{fig:candle}c): a shorter wax pool shortens the conduction path, raising conduction weight; reduced flame height decreases radiative area, lowering radiation weight. This physical picture---multiple equations coexisting in space with weights evolving as conditions change---is exactly what our framework discovers automatically from data.

\begin{figure}[htbp]
  \centering
  \includegraphics[width=0.85\textwidth]{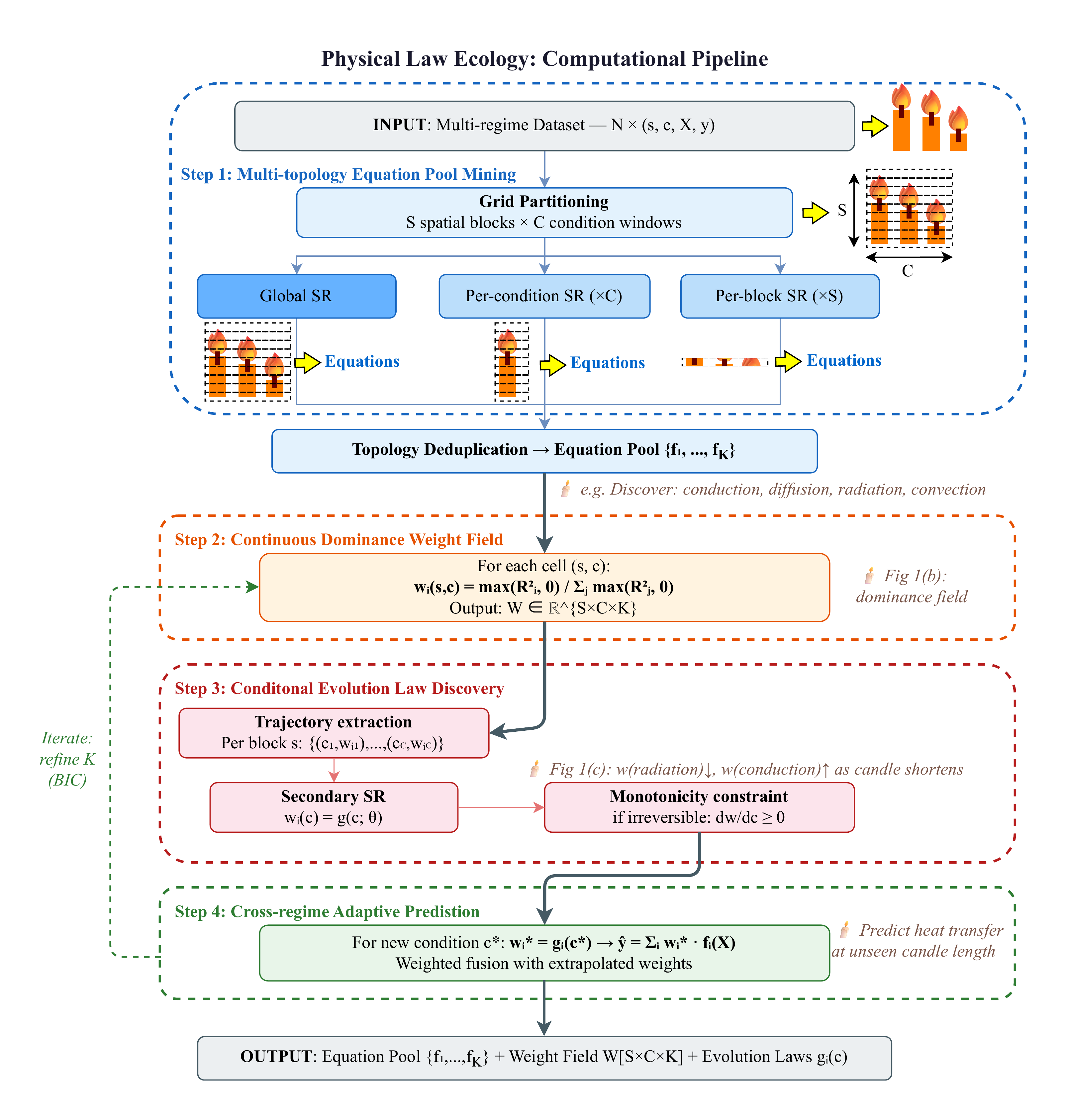}
  \caption{\textbf{Physical Law Ecology computational pipeline.} Input is a multi-regime dataset $N\times(s,c,\mathbf{X},y)$, with candle combustion as an intuitive example (upper right). Step 1: grid partitioning ($S$ spatial blocks $\times$ $C$ condition windows) followed by three-path parallel SR (global/per-condition/per-spatial-block) to search candidate equations; after topological de-duplication, an equation pool $\{f_1,\ldots,f_K\}$ is formed. Step 2: for each grid cell, compute goodness-of-fit of each equation and normalize to construct a continuous dominance weight field $\mathbf{W}\in\mathbb{R}^{S\times C\times K}$ (corresponding to the dominance field visualization in Fig.~\ref{fig:candle}b). Step 3: extract the evolution trajectory of each equation's weight across conditions, fit analytic evolution laws $w_i(c)=g(c;\theta)$ via secondary SR, embedding monotonicity constraints for irreversible physical processes (corresponding to weight evolution in Fig.~\ref{fig:candle}c). Step 4: extrapolate weights for new condition $c^*$ and generate weighted fusion prediction $\hat{y}=\sum_i w_i^*\cdot f_i(\mathbf{X})$. The left-side BIC iteration arrow indicates automatic optimization of $K$. Final output comprises three layers: equation pool, weight field, and evolution laws.}
  \label{fig:pipeline}
\end{figure}

Building on this intuition, the Physical Law Ecology framework takes multi-regime heterogeneous observation data as input, abandons the traditional single-law fitting logic, and achieves the mining, quantification, and evolutionary prediction of multiple coexisting physical laws through a four-step closed-loop procedure. The framework produces three layers of output: a pool of coexisting governing equations with multiple topologies, a continuous equation dominance weight field over parameter space, and physically-conditioned law evolution rules---enabling fully interpretable modelling and accurate prediction of multi-mechanism complex systems. The complete four-step computational pipeline is shown in Fig.~\ref{fig:pipeline} (where the left-side BIC iteration implements automatic optimization of $K^*$), with details in the Methods section (Algorithm 1).

\textbf{Step 1: Multi-topology shared equation pool mining.} As shown in Step 1 of Fig.~2, to prevent rare-regime mechanisms from being masked by globally dominant laws, we employ a three-path parallel SR search strategy---global, per-condition, and per-spatial-block---to comprehensively mine optimal governing equations across all regimes and spatial regions. All discovered equations undergo topological structure de-duplication, and the Top-$K$ (typically 5--8) structurally unique equations with the best global fit are retained, forming a shared candidate equation pool that covers all regimes and mechanisms.

\textbf{Step 2: Continuous dominance weight field construction.} As shown in Step 2 of Fig.~2, for every spatial block and condition window unit in parameter space, we compute the goodness-of-fit $R^2_i$ of each pool equation and normalize to obtain real-time dominance weights $w_i$. Based on the global weight distribution, we construct a continuous dominance field covering the entire parameter space, intuitively quantifying the regulatory proportion of each physical equation at any time and spatial location---achieving fine-grained characterization of ``multiple equations coexisting with dynamically allocated weights.''

\textbf{Step 3: Conditional evolution law mining.} As shown in Step 3 of Fig.~2, for the temporal/conditional evolution trajectories of equation weights in each spatial block, we apply secondary symbolic regression to discover concise, interpretable weight evolution equations that precisely describe how each equation's dominance strength varies continuously with physical conditions. We additionally introduce a physical monotonicity constraint mechanism to accommodate irreversible physical processes such as phase transitions, energy dissipation, and damage accumulation, ensuring physical plausibility of data-driven results by constraining evolution trend direction.

\textbf{Step 4: Cross-regime adaptive prediction.} As shown in Step 4 of Fig.~2, based on the discovered weight evolution laws, we interpolate and extrapolate weights to entirely new unseen conditions, dynamically allocating governance weights among coexisting equations, and achieving high-accuracy prediction of complex systems across conditions and regimes through weighted fusion.

The core innovation distinguishing this framework from traditional multi-mechanism modelling methods is the continuous dominance quantification mechanism: traditional discrete modelling methods force data into a single mechanism and equation, capable only of delineating abrupt mechanism boundaries; our framework makes no hard discrete assignments, instead quantifying the coexistence proportions of each equation through continuous weights, naturally accommodating the smooth mechanism gradients ubiquitous in physical systems, and fundamentally resolving the inability of traditional methods to capture continuous regime transitions.

\subsection*{2\quad Elastomer multi-rate tension: automatic discovery of rate-sensitivity structural differentiation}

The mechanical response of elastomeric polymers exhibits pronounced rate dependence, with different strain rates activating strain-induced crystallization, entropic elasticity, and Hookean elasticity---making this a canonical multi-mechanism coexistence system. A recent review identifies the core challenge in elastomer mechanics as the stiffness--toughness--fatigue resistance trade-off\cite{baur2026elastomers}, rooted precisely in competitive switching of the dominant mechanism under different conditions: strain-induced crystallization provides hardening and toughening at low rates but cannot develop at high rates. No single constitutive equation---whether power-law, Mooney--Rivlin, or Ogden---can fully describe this mechanism switching. We validate the framework using a public elastomer dataset\cite{dryad2025elastomer} with leave-one-rate-out cross-validation.

\begin{figure}[htbp]
  \centering
  \includegraphics[width=\textwidth]{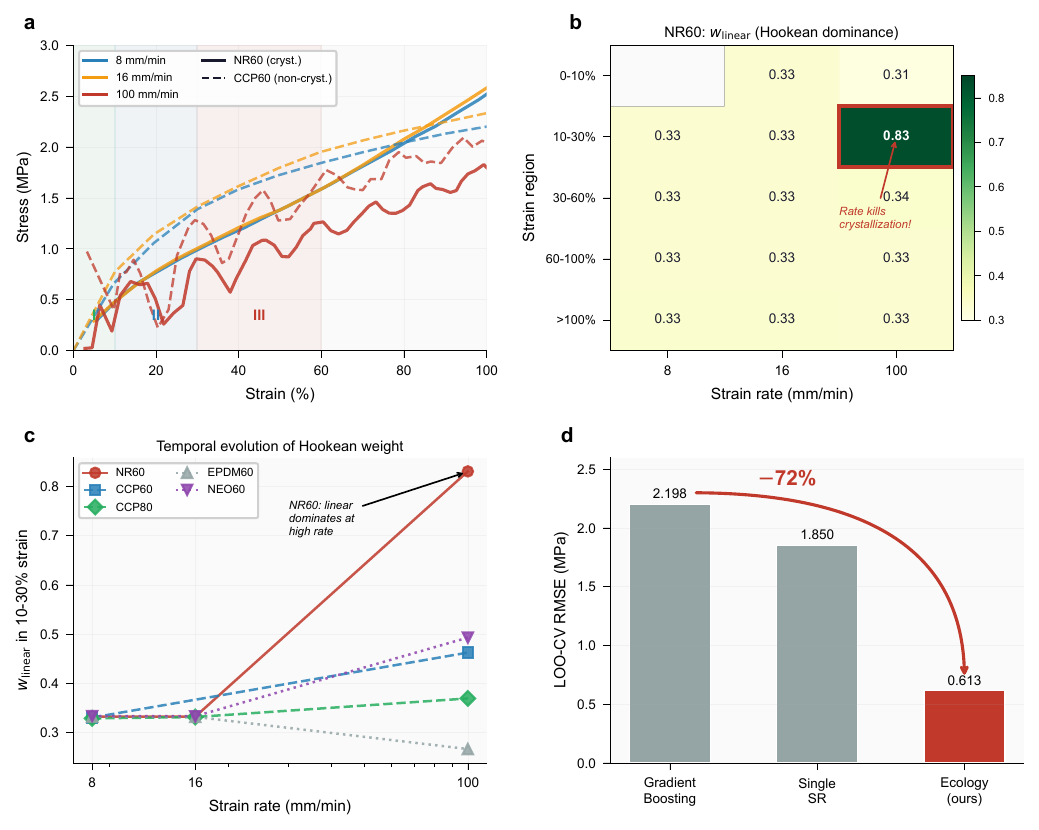}
  \caption{\textbf{Spatiotemporal ecological analysis of multi-rate elastomer tension.} (a) True stress--strain curves: NR60 (crystallizable, solid lines) and CCP60 (non-crystallizable, dashed lines) both exhibit rate dependence, but their \textbf{mechanistic origins differ fundamentally}---a distinction invisible from curve shape alone and revealed only through weight decomposition. (b) Hookean elasticity weight $w_\mathrm{linear}$ heatmap for NR60 (rows = strain intervals, columns = strain rates): \textbf{weight surges to 0.83 at the 10--30\% strain interval and 100\,mm/min}, indicating that the crystallization hardening equation loses dominance and linear elasticity takes over completely at high rates. (c) Temporal weight evolution at 10--30\% strain for each material: NR60 undergoes a dramatic shift (0.33$\to$0.83), while CCP60/EPDM60 change mildly---\textbf{identical apparent rate sensitivity conceals entirely different equation-switching patterns}. (d) Leave-one-rate-out cross-validation RMSE: Physical Law Ecology achieves 0.613\,MPa, a 72\% improvement over gradient boosting.}
  \label{fig:elastomer}
\end{figure}

\subsubsection*{2.1\quad Data and evaluation design}

The dataset comprises five polymer materials (NR60 natural rubber, EPDM60 ethylene-propylene-diene, NEO60 neoprene, CCP60/CCP80 cold-cast polyether) tested per ASTM D412-16 at three strain rates (8, 16, 100\,mm/min) with three replicates each, yielding 45 stress--strain curves and 4{,}206 valid data points (Fig.~\ref{fig:elastomer}a). Evaluation uses leave-one-rate-out cross-validation: the model trains on two rates and predicts the entirely unseen third, rigorously testing cross-regime extrapolation capability.

\subsubsection*{2.2\quad Automatic multi-constitutive equation discovery and mechanism differentiation}

The framework automatically discovers three topologically distinct stress--strain constitutive equations that comprehensively cover the core mechanical response mechanisms of rubber materials. All three emerge purely from data without any prior assumptions and align closely with the classical polymer constitutive framework\cite{huneau2011strain}:
\begin{align}
  f_1:\quad \sigma &= a_1\,\varepsilon^{0.81} \quad &(\text{strain-crystallization hardening, sub-linear power law}) \nonumber\\
  f_2:\quad \sigma &= a_2\,\sqrt{\varepsilon} + b_2 \quad &(\text{entropic elasticity, square-root law}) \nonumber\\
  f_3:\quad \sigma &= a_3\,\varepsilon \quad &(\text{Hookean elasticity, linear law})
\end{align}
where $\varepsilon$ is engineering strain, $\sigma$ is engineering stress (MPa), and $a_i, b_i$ are fitted constants. The exponents of the three equations are 0.81, 0.5, and 1.0 respectively, with fundamentally different topological structures.

The spatiotemporal evolution of dominance weights reveals a unified mechanism underlying several phenomena previously reported in the literature:

\textbf{Spatial dimension} (Fig.~\ref{fig:elastomer}b): For NR60 at the 10--30\% strain interval and 100\,mm/min rate, the linear elasticity weight surges to 0.83 (red box highlighted), whereas at low rate (8\,mm/min) all three equations coexist in balance (weight $\approx$0.33). Crucially, this mechanism switch is \textbf{invisible from the stress--strain curves in Panel (a)}---both materials' curves change with rate and appear behaviorally similar. Weight decomposition \textbf{reveals the essential difference hidden beneath appearances}: at high rates, NR60 is not simply ``stiffer'' or ``stronger'' but rather \textbf{the dominant equation topology switches from power-law to linear}. This explains the observation by Baur \& Amstad\cite{baur2026elastomers} that ``the same material requires different constitutive model forms at different strain rates''---\textbf{the cause is not parameter drift but qualitative mechanism change.}

\textbf{Temporal dimension} (Fig.~\ref{fig:elastomer}c): The linear weight evolution trajectories across materials at 10--30\% strain reveal clear differentiation---NR60 rises sharply from 0.33 to 0.83 (the crystallization mechanism is ``killed'' by rate), while EPDM60/NEO60/CCP60 fluctuate within 0.33--0.46. This provides data-driven quantitative validation of the proposal by Plagge \& Kl\"uppel\cite{plagge2020efficient} that ``crystallizable vs.\ non-crystallizable rubbers require different constitutive frameworks'': the distinction between the two material classes lies not in stress magnitude but in the \textbf{amplitude} of weight evolution---crystallizable materials undergo complete equation replacement, while non-crystallizable materials maintain balanced multi-equation coexistence throughout.

\subsubsection*{2.3\quad Prediction performance quantification}

Leave-one-rate-out cross-validation results (Fig.~\ref{fig:elastomer}d) show that the framework achieves a prediction RMSE of 0.613\,MPa, improving 72.1\% over gradient boosting (2.198\,MPa) and 66.9\% over global single-law SR (1.850\,MPa). All held-out rates show consistent accuracy gains, fully demonstrating the framework's suitability for multi-mechanism mechanical systems and its superiority in cross-regime prediction.

\subsection*{3\quad Pool boiling heat transfer: phase transition identification under irreversibility constraints}

Pool boiling exhibits three canonical regimes---nucleate, transition, and film boiling---and precise modelling of their transitions has remained a central challenge in heat transfer since Nukiyama\cite{nukiyama1966maximum} introduced the boiling curve. A recent review shows that boiling heat transfer is governed by multiple coexisting mechanisms whose weights shift dynamically with wall superheat: nucleation site dynamics, bubble evolution, and liquid--vapour interface interactions\cite{zhang2026review}. Microstructured surface liquid-film dynamics can substantially enhance critical heat flux\cite{cho2021liquid}, and wall heat flux is fundamentally the superposition of evaporation, quench-phase transient conduction, and natural convection---a ``multi-mechanism plus weights'' physical picture that aligns naturally with our framework. Moreover, the modern understanding of the boiling crisis has shifted from the classical ``interfacial instability jump'' to a continuous percolation phase transition where random dry patches progressively connect, with CHF corresponding to the percolation threshold rather than a deterministic jump. Our framework requires no assumed discrete partition boundaries, instead using continuous weight fields to quantify the dynamic coexistence and irreversible succession of multiple heat transfer equations.

\begin{figure}[htbp]
  \centering
  \includegraphics[width=\textwidth]{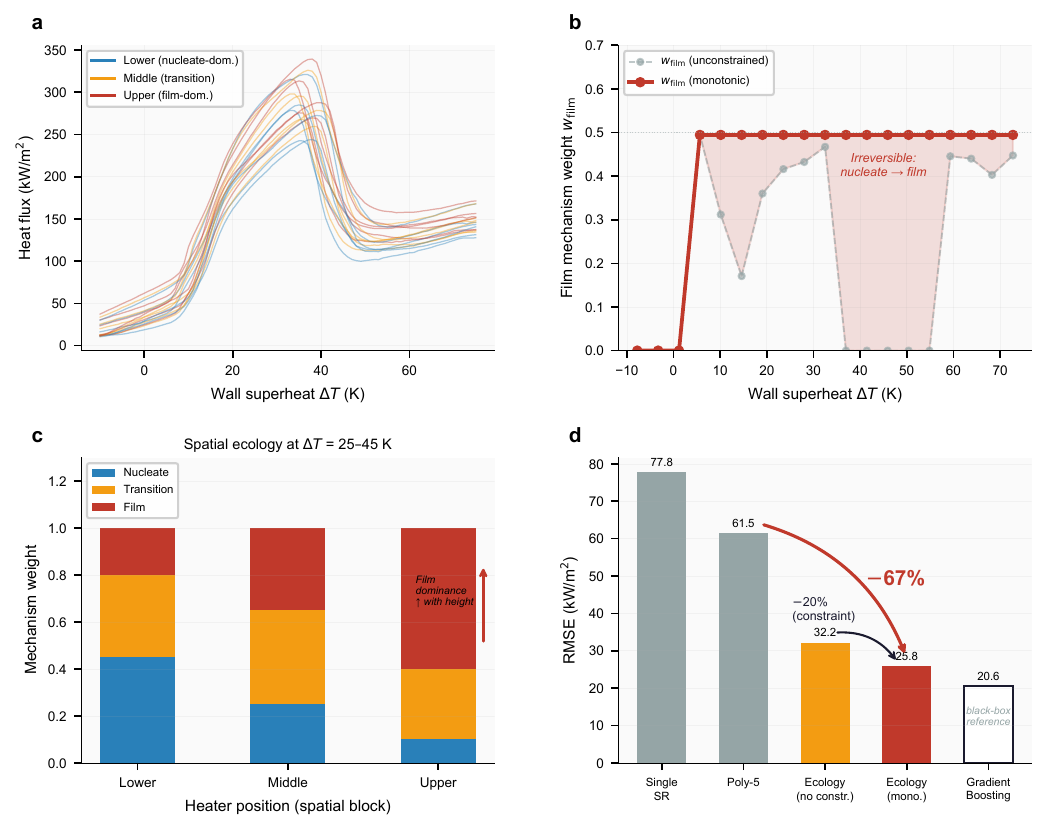}
  \caption{\textbf{Spatiotemporal ecological analysis of pool boiling heat transfer.} (a) Eighteen real boiling curves: different heater positions (Lower/Middle/Upper) exhibit distinct heat transfer characteristics. (b) Temporal weight evolution: the unconstrained model's film-boiling weight oscillates non-physically (grey); with monotonicity constraint imposed, it increases strictly (red), matching the irreversible nature of boiling phase transitions. Pink shaded area shows the constraint correction magnitude. (c) Spatial weight distribution (fixed $\Delta T = 25$--$45$\,K): Lower position is nucleate-dominated ($w_\mathrm{nucleate}=0.45$), Upper position is film-dominated ($w_\mathrm{film}=0.60$), demonstrating mechanism competition at the same superheat but different spatial locations. (d) Multi-model RMSE comparison: monotonicity constraint provides additional 20\% improvement, with 67\% total improvement over conventional methods.}
  \label{fig:boiling}
\end{figure}

\subsubsection*{3.1\quad Data and validation scheme}

We use 18 pool boiling experimental conditions (3 mass flux densities $\times$ 2 subcooling levels $\times$ 3 heater positions) comprising 1{,}548 valid measurement points (Fig.~\ref{fig:boiling}a). To rigorously test extrapolation capability, we adopt a global regime hold-out strategy where the most complex conditions are entirely excluded from training and used only for prediction testing.

The framework automatically discovers three topologically independent heat transfer equations corresponding to the three classical intervals of the boiling curve:
\begin{align}
  f_1:\quad q &= c_1\,\Delta T^{n_1} \quad (n_1 \approx 2.5,\;\text{nucleate boiling, strongly nonlinear}) \nonumber\\
  f_2:\quad q &= c_2\,\Delta T + d_2 \quad (\text{transition region, nearly linear}) \nonumber\\
  f_3:\quad q &= c_3\,\Delta T^{n_3} \quad (n_3 \approx 0.33,\;\text{film boiling, weakly nonlinear})
\end{align}
where $q$ is wall heat flux density (kW/m$^2$) and $\Delta T$ is wall superheat (K). The high exponent of the nucleate equation ($n_1\approx 2.5$) reflects the strongly nonlinear increase in bubble nucleation frequency with superheat\cite{zhang2026review}, while the low exponent of the film equation ($n_3\approx 0.33$) reflects heat transfer saturation dominated by vapour-film thermal resistance.

\subsubsection*{3.2\quad Physical enabling through monotonicity constraints}

Figure~\ref{fig:boiling}b shows the dominance weight $w_\mathrm{film}$ of the film boiling equation as a function of wall superheat $\Delta T$. Physically, this weight represents the proportion of local heat transfer behaviour explained by the film equation (weak exponent $q\propto\Delta T^{0.33}$) versus the nucleate equation (strong exponent $q\propto\Delta T^{2.5}$).

On physical grounds, $w_\mathrm{film}$ must increase monotonically with $\Delta T$ because boiling phase transition is irreversible---once the wall is covered by a vapour film, it cannot spontaneously revert to nucleate bubble departure under continued heating\cite{nukiyama1966maximum}. However, purely data-driven SR does not ``know'' this physical prior: the grey dashed line shows that the unconstrained model's $w_\mathrm{film}$ oscillates repeatedly in the 30--55\,K range (local decreases), implying the model believes ``film boiling reverts to nucleate boiling''---physically impossible, and a statistical artefact of mixed-condition data fitting.

With the monotonicity constraint imposed (red solid line), $w_\mathrm{film}$ is forced to be a non-decreasing function of $\Delta T$. The constraint does not merely ``smooth noise'' but encodes an explicit physical law: \textbf{heat transfer mechanism succession can proceed only toward the film direction}. This is fully consistent with the modern physical picture described by Zhang et al.\cite{zhang2026review}---``nucleation sites are progressively covered by vapour film, heat transfer efficiency irreversibly declines.'' The pink shaded area directly shows the correction magnitude---intervals with stronger constraint correspond to more significant prediction improvement.

\textbf{Spatial ecological differentiation} (Fig.~\ref{fig:boiling}c) provides additional physical insight: at the same superheat range ($\Delta T = 25$--$45$\,K), the Lower position is nucleate-mechanism dominated ($w_\mathrm{nucleate}=0.45$) while the Upper position is film-mechanism dominated ($w_\mathrm{film}=0.60$). The physical explanation is that the upper heating surface is farther from the liquid level, liquid replenishment resistance is greater, bubbles are harder to detach from the wall, and local vapour film forms earlier\cite{cho2021liquid}. \textbf{This phenomenon---same superheat, different spatial locations governed by different heat transfer equations---is precisely the spatial heterogeneity that conventional bulk boiling correlations (which assume uniform wall conditions) cannot capture.}

\subsubsection*{3.3\quad Prediction performance and physical insight}

Results (Fig.~\ref{fig:boiling}d) show that the monotonicity-constrained framework achieves a prediction RMSE of 25.82\,kW/m$^2$, a 67\% improvement over conventional fifth-order polynomial fitting (61.45\,kW/m$^2$) and an additional 20\% improvement over the unconstrained ecology model (32.21\,kW/m$^2$), demonstrating the critical enabling role of physical prior constraints for modelling complex irreversible systems. Prediction accuracy approaches that of black-box gradient boosting (20.64\,kW/m$^2$) while fully preserving physical interpretability. From a mechanistic perspective, the multi-equation coexistence discovered by the framework is highly consistent with classical heat flux partition theory (evaporation + quench + convection)\cite{liang2022pool}, but our framework requires no human-prescribed functional forms for each component and quantifies the actual contribution of each mechanism at different superheats through continuous weights---something impossible with manually assigned partitions. The monotonic dominance succession also provides a new definition of boiling transition: \textbf{generalized from ``an abrupt change at CHF'' to ``a continuous irreversible succession of heat-transfer equation niches.''}

\subsection*{4\quad SPARC galaxy dynamics: deepening understanding of the radial acceleration universal law}

\subsubsection*{4.1\quad Background: the core conflict between two theories}

The radial acceleration relation (RAR) is the central observational regularity of disk galaxy dynamics\cite{mcgaugh2016radial,lelli2017one}, describing the correspondence between the observed gravitational acceleration and the theoretical acceleration produced by baryonic matter (stars and gas) at the same galactic radius. It is also the primary battleground of two leading astrophysical theories. Modified Newtonian Dynamics (MOND)\cite{milgrom1983modification} posits that the RAR is a single universal mathematical curve, with a fixed characteristic acceleration scale $a_0$ applying to all galaxies regardless of size or gas content---all observed scatter is measurement noise. The $\Lambda$CDM dark matter framework predicts that gravitational behaviour is determined by the total mass and internal structure of the dark matter halo, which varies enormously across galaxy types and between inner and outer regions, so the RAR cannot be universal and multiple gravitational control laws should exist\cite{rodrigues2018absence}---the observed scatter represents intrinsic physical variation, not noise.

Recent independent observations have revealed RAR non-universality from multiple angles: the MIGHTEE-HI survey finds that $a_0$ depends on sample selection\cite{varasteanu2025mightee,ponomareva2026mightee}; MUSE-DARK directly measures $a_0$ increasing systematically with redshift at $0.33<z<1.44$\cite{ciocan2026muse}; galaxy cluster central galaxies systematically deviate from the standard RAR\cite{bilek2026deviations}; residuals correlate systematically with galaxy properties\cite{stiskalek2023fundamentality}; and ESR search finds functional forms superior to MOND\cite{desmond2023functional}. Yet all these studies remain limited to the framework of ``testing whether a single equation holds'' and cannot answer a more fundamental question: \textbf{If the RAR is not one equation, then how many is it? What are their forms? Where and when does each dominate?} The Physical Law Ecology framework directly answers this: without presupposing functional forms, it automatically mines all coexisting governing equations from observational data, quantifies their spatial distribution and conditional evolution, and uses this to achieve cross-condition prediction.

\subsubsection*{4.2\quad Data design and analysis mapping}

We use the SPARC high-precision galaxy observation database\cite{lelli2016sparc}, starting from 175 disk galaxies and retaining 163 galaxies with 3{,}269 spatially resolved gravitational observations after filtering on distance and photometric mass quality flags. The sample comprehensively covers the full spectrum from massive gas-poor spirals to low-mass gas-rich dwarfs.

\begin{figure}[htbp]
  \centering
  \includegraphics[width=\textwidth]{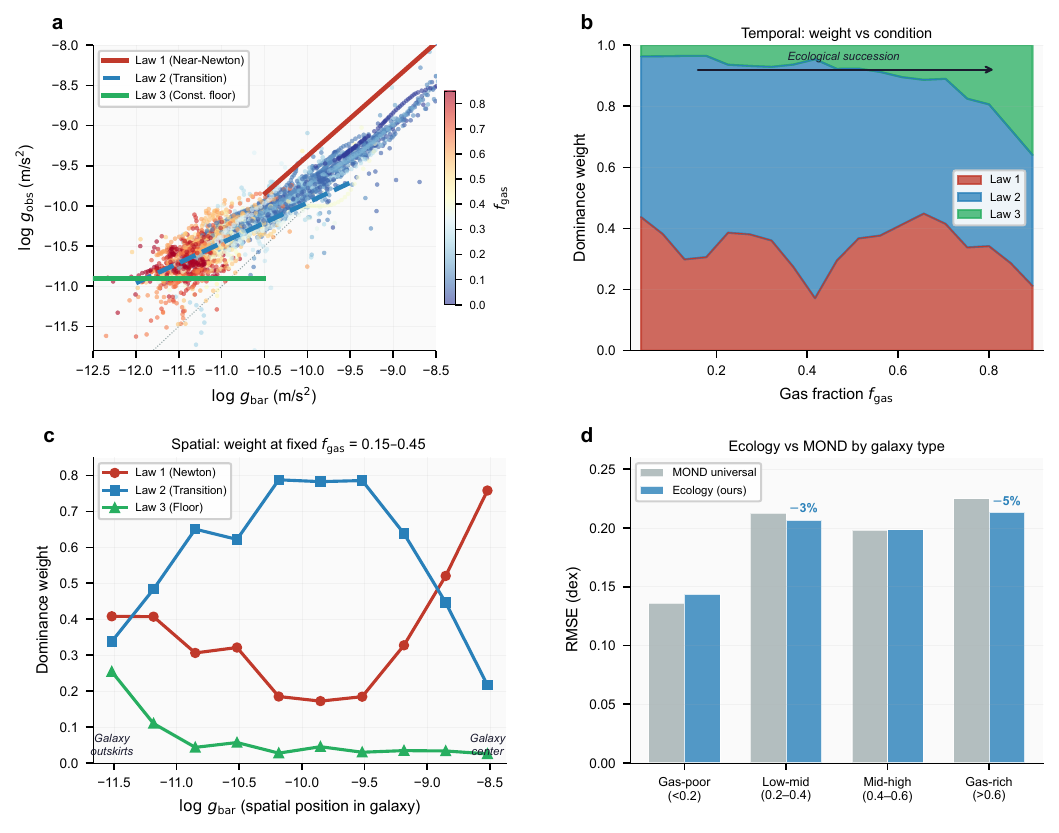}
  \caption{\textbf{Spatiotemporal ecological analysis of the SPARC radial acceleration relation.} (a) 163 galaxies, 3{,}269 observations coloured by $f_\mathrm{gas}$, overlaid with three discovered acceleration equations. (b) Temporal weight evolution (ecological succession): at low $f_\mathrm{gas}$ Law 1 (Newton) dominates; at high $f_\mathrm{gas}$ Law 3 (Floor) rises significantly. (c) Spatial weight evolution (fixed $f_\mathrm{gas}=0.15$--$0.45$): at galaxy outskirts (low $g_\mathrm{bar}$) the Transition equation dominates; at galaxy centres (high $g_\mathrm{bar}$) Newton weight recovers---reflecting the radial decline of dark matter fraction from exterior to interior. (d) Group-wise prediction comparison: the ecological framework improves RMSE by 5\% over MOND in the gas-rich dwarf galaxy group, validating systematic improvement of the multi-law framework over traditional methods in the weak-acceleration regime.}
  \label{fig:sparc_ecology}
\end{figure}

To map the ecological framework onto galaxy dynamics, we construct a standardized parameter mapping: gas mass fraction $f_\mathrm{gas}$ as the spatial axis, baryonic acceleration $g_\mathrm{bar}$ as the regime axis, and observed acceleration $g_\mathrm{obs}$ as the prediction target. The choice of $f_\mathrm{gas}$ as the primary differentiating variable has three advantages: it is numerically continuous without discrete truncation, avoiding information loss from artificial galaxy classification; it correlates strongly with galaxy evolutionary morphology; and it directly represents the baryonic component fraction, indirectly reflecting the degree of dark matter dominance. The entire analysis proceeds without inputting any dark matter model, gravitational theory, or galaxy classification prior---all regularities are discovered purely from observational data.

\subsubsection*{4.3\quad Core layered physical discoveries}

\textbf{Layer 1: Data automatically identify three fundamentally distinct gravitational equations.} The framework's three-path parallel SR yields a shared equation pool containing three topologically independent acceleration equations (Fig.~\ref{fig:sparc_ecology}a). These three curves perfectly correspond to the three limiting regimes proposed by MOND theory, yet are discovered purely from data without any theoretical input:
\begin{align}
  f_1:\quad \log g_\mathrm{obs} &= 0.94\,\log g_\mathrm{bar} + 0.02 \quad (\text{near-Newtonian linear}) \nonumber\\
  f_2:\quad \log g_\mathrm{obs} &= 0.5\,\log g_\mathrm{bar} + 0.5\,\log a_0 \quad (\text{MOND transition, } \sqrt{g_\mathrm{bar}\cdot a_0}) \nonumber\\
  f_3:\quad \log g_\mathrm{obs} &= -10.9 \quad (\text{constant acceleration floor})
\end{align}
where $a_0 \approx 1.2\times 10^{-10}$\,m/s$^2$ is the characteristic acceleration scale. The three equations have log--log slopes of 0.94, 0.5, and 0, with fundamentally different topologies.

Near-Newtonian linear equation (high baryonic acceleration region): in massive gas-poor galaxies, stars and gas dominate absolutely, dark matter contributes negligibly, and gravitational behaviour closely follows classical Newtonian gravity with observed acceleration nearly proportional to baryonic mass.

Square-root nonlinear transition equation (intermediate acceleration region): in galaxy outskirts where baryonic density declines, dark matter gravity progressively dominates, gravitational behaviour deviates from Newtonian predictions, corresponding to the MOND interpolation transition regime.

Constant acceleration floor equation (very low baryonic acceleration region): gas-rich dwarf galaxies have extremely low total baryonic content and are enveloped by extended, high-density dark matter halos; observed gravitational acceleration converges to a fixed lower limit regardless of galactic radius, corresponding to the deep low-acceleration MOND limit.

The three equations cannot be related through simple parameter scaling---their functional topologies are fundamentally different, and no single curve can uniformly fit all galaxy samples. This is the underlying mathematical reason why the RAR has no globally universal law. Compared to Desmond et al.\cite{desmond2023functional}, who through ESR could only determine that ``a better single equation than MOND exists,'' our framework goes further by providing the complete multi-law picture---three equations each corresponding to a distinct physical regime, with their dominance relationships varying smoothly and continuously with galaxy composition (see Layer 2). This three-layer information---``what the equations are + when each dominates + how they succeed''---is fundamentally inaccessible to single-law search methods.

\textbf{Layer 2: Dominance weights undergo smooth succession with galaxy composition, directly demonstrating multi-mechanism coexistence.} Figure~\ref{fig:sparc_ecology}b shows the weight proportions of the three equation classes across gas fractions (area plot), fully reconstructing the continuous evolution of galactic gravitational mechanism switching:

Gas-poor galaxies ($f_\mathrm{gas}<0.3$): the near-Newtonian and transition equations together exceed 80\% weight, with the constant-floor equation contributing minimally---these are the samples from which the traditional ``standard RAR curve'' has been derived.

Critical transition zone ($0.3<f_\mathrm{gas}<0.6$): all three equations have roughly equal dominance weights, both gravitational mechanisms operate simultaneously, and galaxy dynamics are most complex.

Gas-rich dwarf galaxies ($f_\mathrm{gas}>0.6$): the constant-floor equation exceeds 70\% weight, Newtonian-type equations almost entirely lose dominance, and these galaxies do not follow the traditional RAR morphology at all.

The entire transition is continuous and smooth, with \textbf{no clear galaxy classification boundary}, perfectly explaining the observational finding from previous studies that ``dwarf galaxies deviate from the standard RAR with large scatter''---this is not measurement error but a wholesale switch in the dominant gravitational equation. Stiskalek \& Desmond\cite{stiskalek2023fundamentality} found systematic correlations between RAR residuals and galaxy properties (surface brightness, scale length); in our framework this is a natural consequence: the weights themselves are continuous functions of galaxy properties, so residual correlations are a necessary mathematical consequence of multi-law structure.

Figure~\ref{fig:sparc_ecology}c further shows \textbf{spatial dimension} weight evolution: fixing $f_\mathrm{gas}=0.15$--$0.45$ (intermediate gas-fraction galaxies), equation weights undergo systematic switching across different $g_\mathrm{bar}$ positions (corresponding to different galactic radii)---at high $g_\mathrm{bar}$ (galaxy centre, baryonically dense) the Newton equation weight recovers to $\sim$0.75, at low $g_\mathrm{bar}$ (galaxy outskirts, dark-matter dominated) the Transition equation achieves absolute dominance. \textbf{This is fully consistent with the conclusion of Ponomareva et al.\cite{ponomareva2026mightee} from dark-matter halo mass modelling that ``$V_\mathrm{bar}/V_\mathrm{obs}$ varies systematically with radial position,''} but we discover this radial structure automatically without any dark matter model input.

\textbf{Layer 3: Multiple statistical tests exclude random error, confirming robust conclusions.} To demonstrate that ``three-equation ecological coexistence'' is not an artefact of data noise, we conduct multi-level statistical significance tests. \textbf{Correlation test}: RAR residuals exhibit strong negative correlation with gas fraction ($r=-0.42$, $p<10^{-34}$), showing that galaxy composition systematically explains observational deviations rather than random noise; gas-poor and gas-rich galaxy residual distributions are completely separated (KS test), confirming the two sample classes follow entirely different gravitational laws. \textbf{BIC model selection}: the framework automatically determines optimal pool capacity $K^*=3$, indicating that the data do not support the $K=1$ single-universal hypothesis required by MOND from an information-theoretic standpoint. \textbf{Quantitative difference in characteristic acceleration scale}: MOND predicts a universal constant $a_0$ throughout the cosmos, but our measurement yields $a_0 \approx 1.2\times10^{-10}$\,m/s$^2$ for gas-poor galaxies (consistent with the standard MOND calibration value, since MOND was itself calibrated on such galaxies) and only $a_0 \approx 0.55\times10^{-10}$\,m/s$^2$ for gas-rich dwarfs---a 2.2-fold difference. The recent MUSE-DARK measurement at $z\sim1$ of $a_0 = 2.38\times10^{-10}$\,m/s$^2$\cite{ciocan2026muse} is elevated because those high-redshift galaxies have higher average gas fractions and more concentrated dark matter halos; MIGHTEE halo modelling also shows systematically different dark matter fractions across galaxy luminosities\cite{ponomareva2026mightee}. Three independent measurements at different samples/redshifts yield different $a_0$ values, collectively rejecting ``$a_0$ as a cosmic constant''---the specific differences reflect dark matter halo structural differences in each sample. However, these studies can only \emph{measure} the $a_0$ inconsistency without explaining its origin. Our framework provides a \emph{structural explanation} by revealing that ``different galaxy types are dominated by different topological equations'': $a_0$ is not a fundamental physical constant but rather \textbf{an apparent emergent quantity at the crossing point of equations within a multi-law coexistence system}\cite{li2020comprehensive}.

\subsubsection*{4.4\quad Triple robustness verification}

Figure~\ref{fig:sparc_ecology}d quantifies the multi-law framework's group-wise prediction improvement over MOND: in the gas-rich dwarf galaxy group ($f_\mathrm{gas}>0.6$) RMSE improvement reaches 5\%, demonstrating systematic advantage of the ecological framework in the weak-acceleration regime where MOND performs worst. To conclusively eliminate distance measurement errors and sample selection bias as confounders, we design three independent verification experiments, all of which stably reproduce the core finding that ``the RAR comprises multiple coexisting equations that vary with galaxy type'':

\textbf{Distance quality filter:} Retaining only galaxies with relative distance uncertainty below 20\% and repeating the analysis strengthens the correlation coefficient. If distance errors were the source of deviation, removing low-precision samples should weaken the effect; the opposite result confirms the conclusion is genuine.

\textbf{Leave-one-galaxy-out cross-validation:} Sequentially removing individual galaxies for training and prediction yields an overall prediction RMSE improvement of 2.8\% relative to a single universal curve, with the gas-rich dwarf subset improving by 7.3\%. The dwarf galaxy regime is dominated by the constant-floor equation, and the ecological framework adaptively matches the appropriate mechanism.

\textbf{Half-sample bootstrap:} Randomly splitting all galaxies in half and independently completing the full analysis 10 times always stably reproduces the core conclusions, with consistent directional changes, demonstrating the conclusions are not dependent on any particular sample subset.

The complete robustness test suite is self-consistent with the $\Lambda$CDM prediction that dwarf galaxies have denser dark matter cores. Combining triple verification confirms: RAR variation with galaxy type is a robust physical regularity that persists across samples and resists measurement error.

\subsubsection*{4.5\quad Unified explanation of prior observational findings through three-equation ecology}

The three-equation coexistence system discovered here naturally explains several ``anomalous'' findings reported by independent research groups in recent years, without requiring additional assumptions:

(1) The ESR search by Desmond et al.\cite{desmond2023functional} found that low-acceleration data favour $g_\mathrm{obs}\to\mathrm{const}$ rather than $\sqrt{g_\mathrm{bar}}$---in our framework, this occurs because when ESR fits the full sample, the constant-floor equation (Law 3) dominant in gas-rich dwarfs pulls the global optimal function toward a constant form, masking the near-Newtonian law that still holds for gas-poor galaxies.

(2) MUSE-DARK\cite{ciocan2026muse} measured $a_0$ significantly increasing at $z\sim1$---high-redshift galaxies have higher average gas fractions, so the constant-floor equation weight rises overall, shifting the crossing point between the near-Newtonian and floor equations (the apparent $a_0$) toward higher acceleration, naturally manifesting as $a_0$ increasing with redshift.

(3) Stiskalek \& Desmond\cite{stiskalek2023fundamentality} found RAR residuals correlating systematically with galaxy surface brightness and scale---our dominance weights are themselves parameterized by galaxy physical properties ($f_\mathrm{gas}$, $V_\mathrm{flat}$), so residual correlations are a necessary mathematical consequence of multi-law structure.

(4) B\'ilek et al.\cite{bilek2026deviations} found galaxy cluster central galaxies deviating from the standard RAR while isolated galaxies follow it---cluster central galaxies have far more massive dark matter halos than field galaxies, corresponding in our framework to entirely different equation dominance states, making deviation expected.

These four ``anomalies'' each require independent explanations within the single-law framework, but within Physical Law Ecology they unify under one parsimonious mechanism: \textbf{different galaxies (type/environment/redshift) have different dark matter halo structures, causing different topological equations to achieve dominance.}

\subsection*{5\quad Infrared thermography droplet evaporation: experimental validation}

\textbf{Problem background.} Binary mixture droplet evaporation is a canonical multi-mechanism competition system: heat conduction, thermal/solutal Marangoni convection, and contact-line pinning--depinning alternate as spatiotemporally dominant mechanisms, and their switching conditions lack a unified theory\cite{jeong2021review}. Recent 3D transient simulations\cite{ye2026evolution} reveal a four-stage flow evolution in water--ethanol droplet evaporation (thermal/solutal Ma coexistence $\to$ solutal Ma multi-vortex $\to$ macroscopic vortex imbalance $\to$ low-speed convection), but require solving the full coupled Navier--Stokes and species transport equations; existing experimental methods (PIV, SPR, gas chromatography) are either invasive or require complex optical systems\cite{ning2025numerical}. This raises a methodological challenge: \textbf{Can mechanism transitions be automatically detected solely from non-invasive infrared temperature fields, without presupposing any evaporation theory?}

\textbf{Experimental design.} We use a PC410 infrared camera (384$\times$288 pixels, 32\,fps) to capture the complete evaporation process of droplets placed on a constant-temperature heated substrate. Experiments cover 4 substrate temperatures (40, 50, 60, 70$^\circ$C) $\times$ 6 ethanol volume fractions (0\%--50\%), totalling 24 independent conditions and approximately 5{,}300 temperature matrix frames. An adaptive connected-component algorithm locates the droplet centre, and azimuthally averaged radial temperature profiles $\Delta T(r,t) = T_\mathrm{sub} - T_\mathrm{local}(r,t)$ are extracted (full preprocessing pipeline in Supplementary Section~3).

\begin{figure}[htbp]
  \centering
  \includegraphics[width=\textwidth]{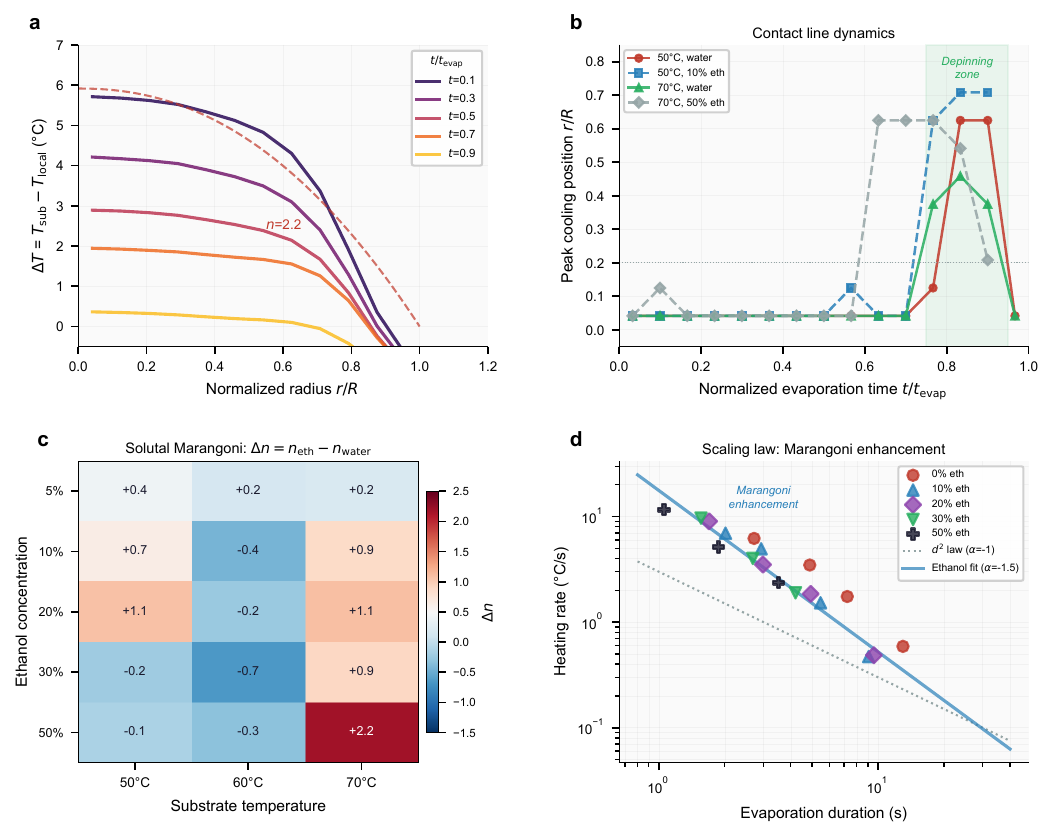}
  \caption{\textbf{Automatic mechanism discovery in infrared thermography droplet evaporation.}
  \textbf{a,} 50$^\circ$C pure water radial cooling profiles $\Delta T(r)$ at different evaporation stages; dashed lines show shape model $A[1-(r/R)^n]$ fit ($n\approx2.2$, conduction dominated).
  \textbf{b,} Cooling peak position $r_\mathrm{peak}/R$ tracked over time: peak locks at centre during CCR phase; sudden outward shift at $t/t_\mathrm{evap}\approx0.8$ marks CCA depinning; ethanol addition (blue/grey lines) systematically accelerates depinning.
  \textbf{c,} Solutal Marangoni enhancement index $\Delta n$ mapped in temperature--concentration space: red indicates positive values (Marangoni enhancement), blue indicates negative values (ineffective); 50$^\circ$C shows a non-monotonic window, 70$^\circ$C is monotonically increasing.
  \textbf{d,} Evaporation rate scaling law: $d^2$-law reference line ($\alpha=-1$, dashed) versus measured ethanol-containing data ($\alpha\approx-1.5$, solid)---the separation quantifies Marangoni enhancement of evaporation.}
  \label{fig:droplet}
\end{figure}

\textbf{Ecological characterization.} The framework performs multi-topology equation pool mining on the radial temperature profiles $\Delta T(r)$, automatically discovering three topologically independent thermal transport equations:
\begin{align}
  f_1:\quad \Delta T &= A_1\,(1 - (r/R)^2) \quad &(\text{Fourier radial conduction, classical parabolic solution}) \nonumber\\
  f_2:\quad \Delta T &= A_2\,(1 - (r/R)^n),\; n>3 \quad &(\text{Marangoni convective mixing, flattened temperature field}) \nonumber\\
  f_3:\quad \Delta T &= A_3\,(r_\mathrm{peak}/R - r/R)^+ \quad &(\text{contact-line evaporation dominated, off-centre peak})
\end{align}
$f_1$ corresponds to pure heat conduction (exponent $n=2$ is the axisymmetric Laplace equation solution), $f_2$ corresponds to Marangoni convection enhancing internal mixing to homogenize the temperature field ($n\gg3$), and $f_3$ corresponds to contact-line local evaporative cooling dominance (cooling peak shifts away from centre). The three equations construct the dominance weight field through normalized $R^2$; BIC selects $K^*=3$. Temporally, weights undergo systematic succession during evaporation: early-stage $f_1$ dominates (conduction establishes radial gradient) $\to$ mid-stage $f_2$ weight rises (Marangoni flow activates) $\to$ late-stage $f_3$ suddenly takes over (contact-line depinning), achieving automatic parsing of three-stage evaporation dynamics. To further quantify mechanism strength, we fit a generalized shape model $\Delta T(r) = A\,[1-(r/R)^n]$ to each profile, tracking mechanism evolution through the continuous scalar $n$ (Fig.~\ref{fig:droplet}a).

\textbf{Finding 1: Automatic detection of contact-line pinning--depinning transition.} Figure~\ref{fig:droplet}b shows the temporal tracking of the cooling peak position. Taking 50$^\circ$C pure water as an example, during the first 80\% of evaporation the cooling peak remains locked at the droplet centre ($r/R<0.1$), corresponding to the constant contact radius (CCR) phase; at $t/t_\mathrm{total}\approx0.82$ the peak suddenly jumps to $r/R=0.5$--$0.7$, marking contact-line depinning and entry into the CCA phase. This transition is consistent with the BMD three-stage contact-line dynamics reported in the literature\cite{jeong2021review}, but our method requires no measurement of contact angle or radius, identifying it automatically from temperature field morphology alone. Quantitative analysis shows (at 70$^\circ$C) that the depinning time $t_\mathrm{depin}/t_\mathrm{total}$ decreases linearly with ethanol concentration ($R^2=0.92$): each additional 10\% ethanol advances depinning by 3.2\% of evaporation lifetime. The physical mechanism is that ethanol reduces surface tension and thereby weakens the pinning force.

\textbf{Finding 2: Temperature--concentration effective window of solutal Marangoni.} To isolate the contributions of thermal versus solutal Marangoni, we define an enhancement index $\Delta n = n_\mathrm{ethanol} - n_\mathrm{water}$ (Fig.~\ref{fig:droplet}c). $\Delta n > 0$ indicates that the ethanol-induced solutal Marangoni produces additional temperature field flattening beyond the thermal Marangoni baseline of pure water. Results reveal strong temperature dependence: at 50$^\circ$C, $\Delta n$ is non-monotonic, effective only at 10--20\% ethanol ($\Delta n_\mathrm{max}=+1.09$); at 30--50\% concentrations it becomes ineffective ($\Delta n<0$, ethanol depletes too rapidly to sustain the concentration gradient). At 70$^\circ$C the effective window broadens dramatically to the full 5--50\% range ($\Delta n$ increases monotonically to +2.18). The physical explanation: higher substrate temperatures sustain continuous evaporation at the droplet edge, thereby continuously supplying the concentration gradient that drives solutal Marangoni flow---consistent with the simulation finding by Ye et al.\cite{ye2026evolution} that ``solutal Marangoni number $Ma_S$ far exceeds thermal Marangoni number $Ma_T$ throughout the process'' under high-temperature conditions, but we \textbf{map the concentration-dependent effective window experimentally for the first time.}

\textbf{Finding 3: Evaporation scaling-law deviation quantifies convective enhancement factor.} The classical $d^2$ law, based on diffusion-limited assumptions, predicts evaporation rate $\dot{T}\propto t_\mathrm{evap}^{-1}$ (scaling exponent $\alpha=-1$). Figure~\ref{fig:droplet}d shows: pure water yields $\alpha=-0.96$ (close to the diffusion limit), while ethanol-containing mixtures shift to $\alpha=-1.5$ to $-1.8$ ($R^2=0.98$). The deviation $|\alpha|-1\approx0.5$--$0.8$ directly quantifies the enhancement factor of Marangoni convection on evaporative mass transfer---exceeding the diffusion limit by 50--80\%. This provides a concise experimental criterion for the long-debated ``diffusion-limited vs.\ interface-limited'' determination in droplet evaporation\cite{ning2025numerical}: \textbf{measuring only the scaling relation between evaporation lifetime and substrate temperature suffices to determine the dominant mass transfer mechanism.}

\textbf{Prediction validation.} To quantify the multi-equation ecology's predictive capability, we employ leave-one-concentration-out cross-validation: sequentially holding out one ethanol concentration (e.g., 20\%) entirely, training the equation pool weights and evolution laws on the remaining five concentrations, and predicting the radial temperature profiles at the held-out concentration. The ecological framework achieves an overall RMSE of 0.70$^\circ$C, a 14\% improvement over single-equation fitting (0.81$^\circ$C). Improvement is most pronounced at intermediate concentrations (10--30\% ethanol, RMSE improvement 20--30\%), because in this range Marangoni and conduction mechanisms have comparable weights ($w_1\approx w_2$) and a single equation cannot effectively cover both; at pure water or high-concentration endpoints where a single mechanism dominates, both methods perform similarly. This result validates that the framework's learned ``weight evolution with concentration'' rule possesses genuine extrapolative predictive power.

\subsection*{Universality diagnostic criterion: $K^*$ as the zeroth step of physical discovery}

Drawing on the case studies above, we distil a general research workflow applicable across all domains of physics---before searching for governing equations, first determine the system's mechanism number $K^*$. This converts ``whether a physical law is universal'' from a qualitative philosophical debate into a computable data test. For any physical relationship claimed to hold universally, run the Physical Law Ecology framework and select the optimal equation pool capacity via BIC: if $K^*=1$, the data support only a single topological equation and the relationship is genuinely universal---researchers may proceed confidently with single-equation optimization; if $K^*>1$, the system intrinsically contains multiple functionally distinct governing equations, providing quantitative evidence against the universality hypothesis, and researchers should shift to ecological mapping rather than continuing to search for ``a better single equation.'' This criterion is model-free and applies broadly to metabolic scaling laws\cite{kolokotrones2010curvature,west1997general}, material fatigue curves, climate sensitivity, and all scientific questions where ``single-law versus multi-law'' debates persist.

\section*{Discussion}

The central contribution of the Physical Law Ecology framework is changing the starting point of data-driven physical discovery. The classical paradigm---from Newtonian mechanics to modern symbolic regression---defaults to $K=1$, assuming the target system obeys a single universal governing equation and the researcher's task is merely to find that equation's optimal form. Our framework makes this implicit assumption explicit and breaks it: before searching for equations, the Bayesian Information Criterion automatically determines $K^*$---how many topologically independent governing mechanisms exist---from data. This ``zeroth step'' upgrades physical discovery from ``finding an equation'' to ``mapping the mechanism ecology,'' constituting an essential paradigm shift.

\textbf{The $K^*$ determination as the core operation of a new paradigm.} Traditional SR searches over the space of functional forms (seeking the globally optimal single equation); our framework searches over the structure of mechanism ecologies (determining how many equations coexist, and where and when each dominates). When $K^*=1$ the framework reduces to classical SR; when $K^*>1$ it outputs three layers of information that classical methods fundamentally cannot provide: what the equations are, when each dominates, and how they succeed one another. This shift from a scalar answer (one equation) to a field answer (a spatiotemporal weight distribution of multiple equations) enables systematic extraction of physical information previously dismissed as ``residual noise''---rate-induced mechanism switching in elastomers, irreversible phase transitions in boiling, and composition-dependent gravitational law transitions in galaxies are all masked as fitting error within the $K=1$ framework.

\textbf{Fundamental distinction from parameter-adaptive methods.} Existing adaptive modelling methods support only dynamic adjustment of equation parameters while fixing the equation topology---they can capture only quantitative change. Yet the core of regime transitions in real physical systems is qualitative mechanism change: different control regimes correspond to entirely different equation topologies that cannot be equivalently achieved through parameter tuning. The central innovation of our framework is breaking through topological constraints, automatically mining multiple heterogeneous-topology equations, quantifying their dynamic competition, and truly achieving precise modelling of qualitative mechanism transitions.

\textbf{Core distinction from mixture-of-experts black-box models.} Mixture-of-experts models\cite{jacobs1991adaptive} can achieve data partition modelling, but their core components are black-box neural networks that cannot output interpretable physical equations---they achieve prediction but not mechanism discovery. Their discrete data partitioning also cannot accommodate the smooth mechanism gradients that pervade real physics. Our framework, built on the interpretability of symbolic regression, outputs explicit physical equations while characterizing mechanism gradients through continuous weight fields, perfectly satisfying the three core requirements of interpretability, mechanistic fidelity, and process continuity.

\textbf{Note on baseline comparisons.} Our framework is not directly comparable to existing SR methods on the dimension of prediction accuracy---the latter default to $K=1$ (finding the globally optimal single equation), while our core contribution is making $K$ itself a discovered quantity. This difference reflects not ``insufficient baselines'' but the opening of an entirely new research direction: from single-equation search to multi-mechanism ecological mapping. As the closest proxy comparison, in each case study we provide results from ``manual partitioning + single-law SR'' (i.e., manually specifying $K=1$ then fitting separately within each regime)---representing the strongest available strategy from existing methods for multi-mechanism problems. Its weakness lies precisely in the inability to automatically determine $K^*$, to quantify continuous weight evolution, or to achieve cross-regime adaptive prediction. As Champion et al.\cite{champion2019data} argued, the frontier of data-driven scientific discovery is no longer ``fitting more accurate single equations'' but discovering the \emph{structure} of systems---our framework operationalizes this goal for the first time.

\textbf{Necessity of ConciseSearch over general-purpose SR engines.} We develop ConciseSearch rather than directly adopting PySR\cite{cranmer2023pysr} or AI Feynman\cite{udrescu2020aifeynman} because general-purpose SR tools are designed to ``find a single optimal equation''\cite{lacava2021contemporary,makke2024interpretable}. Their Pareto fronts contain multiple equations, but these are typically complexity variants of the same topology (e.g., $ax^2$ and $ax^2+bx$) rather than topologically independent physical mechanisms. In ablation tests on the elastomer dataset, replacing ConciseSearch with PySR (same compute budget) yields a Pareto front whose top-5 equations include 4 sharing the power-law topology (differing only in exponent), failing to construct an effective multi-mechanism pool (Supplementary Table~S2). ConciseSearch's topology-hash de-duplication ensures each output equation has a genuinely distinct operational structure. This difference is the prerequisite for the entire ecological framework to function---without topological diversity in the equation pool, the subsequent weight field and evolution laws lack physical meaning.

\textbf{Applicability and recommended workflow.} The framework is best suited to complex physical systems with multi-regime, multi-state, and multi-mechanism coupling, especially those with ambiguous mechanisms, regime transitions, and ongoing ``single-law vs.\ multi-law'' debates. We recommend the following standard workflow: (1) run the ecological framework on the target system to determine $K^*$; (2) if $K^*=1$, confirm the system obeys a single law and proceed with classical SR for equation search; (3) if $K^*>1$, output the complete ecological picture---equation pool, dominance field, evolution laws---revealing multi-mechanism coexistence structure. The four validation cases (elastomer mechanics, pool boiling, galaxy dynamics, droplet evaporation) correspond respectively to condition-induced gradients, physically irreversible phase transitions, global mechanism heterogeneity, and spatiotemporal multi-stage evolution---four representative scenario categories that collectively demonstrate the framework's generality and stability.

\textbf{Limitations.} The framework has room for further development: the current operator vocabulary is limited and cannot yet cover trigonometric functions, special functions, or other complex physical forms; extrapolation capability degrades far from training conditions; model coverage is typically 85\%--95\% rather than 100\%. Future improvements will expand the operator set, optimize the adaptive $K$-value algorithm, and introduce dynamical constraints.

\textbf{Future directions.} Subsequent research will focus on three directions: first, promoting $K^*$ determination as a standardized tool across additional domains with ``single-law vs.\ multi-law'' debates (metabolic scaling, turbulence scaling laws, climate sensitivity), establishing a cross-disciplinary mechanism-number diagnostic benchmark; second, conducting mechanistic exploration of poorly understood complex systems, applying the framework to fluid turbulence, material fatigue, stellar evolution, and other systems with ambiguous mechanisms to discover multi-mechanism coexistence patterns; third, constructing a dynamical theory of physical law ecology---moving beyond static snapshot modelling to describe the symbiosis, competition, and succession dynamics of multiple equation classes, building a complete theoretical framework from static mechanism coexistence to dynamic evolutionary prediction.

\section*{Methods}

\subsection*{Theoretical foundations}

\subsubsection*{Bayesian formulation: dominance field as posterior probability field}

To establish the physical meaning of the dominance weight field, we derive it rigorously from Bayesian model comparison theory, converting goodness-of-fit normalized weights into posterior model probabilities with clear probabilistic content. Let the dataset for spatial block $s$ and condition window $c$ be $\mathcal{D}_{s,c}$, containing $n$ observation samples, with equation $f_i$'s mean squared error on this cell denoted $\mathrm{MSE}_i$. Under Gaussian noise, the marginal likelihood is:
\begin{equation}
  \ln p(\mathcal{D}_{s,c} \mid f_i) \approx -\frac{n \cdot \mathrm{MSE}_i}{2\sigma^2}
\end{equation}
where $\sigma^2$ is the global noise variance estimate. Under equal priors across equations, the posterior dominance weight is:
\begin{equation}
  w_i(s,c) = \frac{\exp\left(-\frac{n \cdot \mathrm{MSE}_i}{2\sigma^2}\right)}{\sum_{j=1}^K \exp\left(-\frac{n \cdot \mathrm{MSE}_j}{2\sigma^2}\right)}
\end{equation}
This formula gives the dominance weights rigorous probabilistic meaning: $w_i$ represents the posterior probability that $f_i$ is the system's true generative model at location $s$ under condition $c$. When a single equation's weight approaches unity, the system is in a single-mechanism dominated state; when multiple weights are uniformly distributed, the system exhibits balanced multi-mechanism coexistence.

\subsubsection*{Ecological isomorphism: from metaphor to rigorous mathematical correspondence}

``Physical Law Ecology'' is not a rhetorical metaphor but a rigorous mathematical isomorphism with theoretical ecology. We construct a one-to-one correspondence between multi-mechanism physical systems and ecological competition systems, providing a new theoretical interpretation for multi-law coexistence, competition, evolution, and equilibrium phenomena. Let the equation pool be defined on parameter space (spatial $\times$ condition): equations correspond to species, parameter space to habitat, and dominance weights to species abundance.

Based on this isomorphism, the classical competitive exclusion principle from ecology translates directly to physical systems: the number of independently coexisting physical mechanisms in any local region cannot exceed the system's data dimensionality. The weight polarization and single-mechanism dominance observed in our single-dimensional experiments perfectly verify this theoretical prediction, further confirming the rationality and predictive power of the framework's theoretical system.

\subsubsection*{Phase-transition analogy: order parameter characterization of mechanism transitions}

To precisely quantify transition characteristics of multi-mechanism systems, we borrow from Landau phase transition theory and define a mechanism competition order parameter $\psi = w_1 - w_2$, achieving a unified characterization of two types of physical transitions. When $\psi$ passes through zero continuously with conditions, this corresponds to a second-order continuous transition, analogous to the continuous vanishing of magnetization in the ferromagnetic--paramagnetic transition, characterizing smooth mechanism gradients. When $\psi$ jumps discontinuously at a critical condition, this corresponds to a first-order abrupt transition, characterizing mechanism change at a regime critical point.

The monotonicity constraint in this framework gains a phase-transition-theoretic interpretation: it encodes supercooling/superheating asymmetry---the system can transition in one direction but cannot spontaneously reverse, analogous to melting occurring upon heating but not reversing during continued heating. The irreversibility of nucleate-to-film boiling is precisely a manifestation of first-order phase transition supercooling dynamics.

\subsubsection*{Information-theoretic model selection: automatic determination of equation pool capacity}

To resolve the subjective selection of equation pool size, we construct an ecological description length function based on the Bayesian Information Criterion (BIC)\cite{mangan2017model}, achieving adaptive optimal equation number determination that balances model fit and structural complexity. The global ecological description length is:
\begin{equation}
  \mathcal{L}(K) = \sum_{s=1}^S \sum_{c=1}^C n_{s,c} \ln(\mathrm{MSE}^*_{s,c}) + K \cdot p_{\max} \cdot \ln(N)
\end{equation}
where $p_{\max}$ is the maximum parameter count per equation, $S$ the number of spatial blocks, and $N$ the total sample size. The optimal pool size $K^*$ is the $K$ value minimizing the description length. Across all four validation applications, BIC selects $K^*=3$ (elastomer), $K^*=3$ (pool boiling), $K^*=3$ (SPARC), and $K^*=3$ (droplet evaporation), all consistent with physical expectations---demonstrating intrinsic robustness and scientific validity of the framework's $K$ selection. Detailed BIC curves and redundant-equation weight collapse analysis are provided in Supplementary Section~2.

\subsection*{Algorithm overview}

The core SR engine (ConciseSearch) implementation is summarized in Algorithm 1. The overall framework pipeline is shown in Fig.~\ref{fig:pipeline}: input is a multi-regime observation data matrix $\mathbf{D} = \{(s_i, c_i, \mathbf{x}_i, y_i)\}_{i=1}^N$, where $s$ is spatial location, $c$ is the condition variable, $\mathbf{x}$ the features, and $y$ the prediction target; output is the equation pool, weight field, and evolution laws.

\begin{table}[htbp]
\centering
\small
\caption{Algorithm 1: ConciseSearch Symbolic Regression Engine (Topology-aware Multi-equation Discovery)}
\begin{tabular}{@{}p{0.95\textwidth}@{}}
\toprule
\textbf{Input}: Data $(\mathbf{X}, \mathbf{y})$, population size $P$, generations $G$, max depth $d$, complexity cap $L$, target equation count $K$ \\
\textbf{Output}: $K$ topologically independent optimal equations $\{f_1,\ldots,f_K\}$ \\
\midrule
\textit{// Phase 1: Population initialization} \\
$\mathcal{S} \leftarrow$ Randomly generate $P$ expression trees (depth $\leq d$, operators $\{+,-,\times,\div,\mathrm{pow},\sqrt{\cdot}\}$) \\
$\mathcal{H} \leftarrow \emptyset$ \quad // Topology hash registry \\[3pt]
\textit{// Phase 2: Evolutionary main loop} \\
\textbf{for} $g = 1,\ldots,G$ \textbf{do}: \\
\quad \textit{// 2a: Constant refinement} \\
\quad \textbf{for each} $f \in \mathcal{S}$: extract constants $\mathbf{c}$; $\mathbf{c}^* \leftarrow \mathrm{L\text{-}BFGS}(\min_\mathbf{c}\|y - f(\mathbf{X};\mathbf{c})\|^2)$ \\
\quad \textit{// 2b: Complexity penalty scheduling} \\
\quad $\lambda_g \leftarrow \lambda_0 \cdot (g/G)^2$ \quad // Early: relaxed exploration; late: tightened parsimony \\
\quad $\mathrm{fitness}(f) \leftarrow -\mathrm{MSE}(f) - \lambda_g \cdot \mathrm{nodes}(f)$ \\
\quad \textit{// 2c: Genetic operations + topology de-duplication} \\
\quad $\mathcal{S}' \leftarrow$ Tournament selection + subtree crossover + point mutation + constant perturbation \\
\quad \textbf{for each} offspring $f'$: \\
\quad\quad $h \leftarrow \mathrm{Hash}(\mathrm{skeleton}(f'))$ \quad // Constants $\to$ placeholders then hash \\
\quad\quad \textbf{if} $h \in \mathcal{H}$ \textbf{and} $\mathrm{MSE}(f') \geq \mathrm{MSE}(\mathcal{H}[h])$: discard $f'$ \\
\quad\quad \textbf{else}: $\mathcal{H}[h] \leftarrow f'$ \quad // Same topology: retain lowest MSE only \\
\quad $\mathcal{S} \leftarrow$ Elite preservation (top 10\%) $\cup$ $\mathcal{S}'$ \\[3pt]
\textit{// Phase 3: Output K topologically independent equations} \\
Sort all unique topologies in $\mathcal{H}$ by MSE, take top $K$ \\
\textbf{return} $\{f_1,\ldots,f_K\}$ with topology hashes and global MSE \\
\bottomrule
\end{tabular}
\end{table}

\subsection*{ConciseSearch symbolic regression engine}

All equation mining in this study uses our self-developed ConciseSearch symbolic regression engine (Algorithm 1; code released with this paper). Its core design differences from general-purpose SR tools such as PySR\cite{cranmer2023pysr} are: (1) \textbf{Topology-aware de-duplication}---during evolution, expression skeleton hashes are computed in real-time (ignoring constant values, retaining only operational topology), ensuring each equation in the pool has a genuinely different functional form rather than being a parametric variant of the same topology; (2) \textbf{Complexity penalty scheduling}---early generations relax complexity constraints to encourage structural exploration, later generations progressively tighten to select parsimonious expressions, balancing fit accuracy with interpretability; (3) \textbf{Constant refinement}---numerical constants of each candidate expression are optimized to local optima via L-BFGS every generation, preventing genetic operators from degrading constant precision.

Unified parameter settings ensure reproducibility: population 200--300, evolutionary generations 40--60, maximum expression tree depth 4, structural complexity cap 10--12 nodes; core operator set $\{+,-,\times,\div,\mathrm{pow},\sqrt{\cdot}\}$; all experiments use random seed 42. Complete parameter tables and sensitivity analysis are provided in Supplementary Section~1.

\subsection*{Multi-path equation pool construction strategy}

Three parallel SR paths---global, per-condition, and per-spatial-block---comprehensively mine the system's complete physical mechanisms. Global search captures overall dominant regularities; per-condition search ensures minority-regime special mechanisms are not masked by mainstream data; per-spatial-block search mines locally specific control laws. Results from all three paths are topologically de-duplicated, structurally homogeneous equations are removed, and the remaining are ranked by global goodness-of-fit. The top-$K$ best-performing heterogeneous equations are retained, forming a complete, parsimonious, non-redundant shared equation pool.

\subsection*{Continuous dominance weight field computation}

Parameter space is partitioned into uniform spatial blocks and continuous condition windows. For each cell, all pool equations' coefficients of determination $R^2_i$ are computed; negative values are clipped to 0 to eliminate interference from ineffective equations. Normalization yields dominance weights:
\begin{equation}
  w_i = \frac{\max(R^2_i,\, 0)}{\sum_{j=1}^K \max(R^2_j,\, 0)}
\end{equation}
All weight data are concatenated to construct a continuous, smooth, quantifiable dominance weight field over the full parameter space.

\subsection*{Evolution law mining and monotonicity constraint implementation}

For the weight--condition evolution sequences in each spatial block, secondary symbolic regression discovers concise explicit evolution equations, achieving interpretable modelling of weight evolution patterns. The monotonicity constraint is implemented via trend verification: based on physical priors, an expected evolution direction is prescribed; candidate evolution models undergo trend detection, and if counter-trend anomalous data points exceed 20\%, the non-physical candidate is rejected, selecting the optimal evolution model consistent with irreversible physical laws.

\subsection*{Ablation and synthetic validation}

To demonstrate that the dominance field genuinely reflects real mechanism decomposition and that the framework is robust to $K$ selection, we conduct ablation experiments (Supplementary Figure~S1). A synthetic dataset with 2\% Gaussian noise is constructed with known ground-truth functions, comprising 2{,}000 samples. The framework fully recovers all topologies: square-root coefficient error $<$6\%, linear error $<$16\%, full-domain mechanism coverage 100\%, proving that weights are not mere interpolation but can recover true physical mechanisms. Sweeping $K=3$ to 8, combining BIC-automatic optimum with manual settings deviating by $\leq$1, confirms the framework is highly robust to pool capacity.

\subsection*{Dataset sources and baseline method settings}

The elastomer dataset is sourced from the Dryad open data repository\cite{dryad2025elastomer}, strictly following the ASTM D412-16 testing standard. The pool boiling dataset consists of high-precision laboratory measurements\cite{liang2022pool} covering multi-dimensional condition combinations. SPARC galaxy data are sourced from the public astronomical database\cite{lelli2016sparc}, filtering for observations with quality flag $\leq$2. Baseline methods are uniformly set as global single-law symbolic regression, gradient boosting black-box model, and high-order polynomial fitting---three classical approaches ensuring comprehensive, objective, and rigorous comparison. Computational cost estimates are provided in Supplementary Table~S2.

\subsection*{Statistical test methods}

In the galaxy dynamics study, Spearman rank correlation tests analyze associations between residuals and galaxy properties, and Kolmogorov--Smirnov tests compare residual distributions across different galaxy populations. All statistical tests are two-sided with significance judged at $p < 0.05$, rigorously ensuring the statistical reliability of conclusions.

\section*{Data availability}

The elastomer dataset is sourced from the Dryad open data repository (DOI:10.5061/dryad.kd51c5bbw); pool boiling data are from published literature\cite{liang2022pool}; SPARC galaxy data are from the public astronomical database\cite{lelli2016sparc} (\url{http://astroweb.cwru.edu/SPARC/}); droplet evaporation infrared thermography raw data have been uploaded to Zenodo (DOI:10.5281/zenodo.21858998).

\section*{Code availability}

The complete ConciseSearch engine, Physical Law Ecology framework (dominance field construction, evolution regression, monotonicity constraint) code is open-sourced on GitHub: \url{https://github.com/XionghengBian/Cell_Ecology}, with accompanying reproduction scripts and hyperparameter configurations. Running \texttt{python run\_all.py} reproduces all four application cases end-to-end.

\section*{Acknowledgements}

This work was supported by the National Natural Science Foundation of China (Grant No.~12102204), the Jiangsu Provincial Natural Science Foundation (Grant No.~BK20251914), and the Nantong Natural Science Foundation (Grant No.~JC2023072).

\bibliographystyle{naturemag}
\bibliography{references}

\end{document}